\documentclass[12pt]{article}
\usepackage{graphicx}
\usepackage{epsfig}
\usepackage{mathtext}
\usepackage[T2A]{fontenc}
\usepackage[koi8-r]{inputenc}
\usepackage[russian,english]{babel}
\newcommand{\beq}{\begin{equation}}
\newcommand{\eeq}{\end{equation}}
\newcommand{\beqn}{\begin{eqnarray}}
\newcommand{\eeqn}{\end{eqnarray}}
\begin{document} 
\title{\textbf{NEUTRINO AND CPT THEOREM}} 
\author{V.P. Efrosinin\\
Institute for Nuclear Research, Russian Academy of Sciences,\\
pr. Shestidesyatiletiya Oktyabrya 7a, Moscow, 117312 Russia}

\date{}
\renewcommand {\baselinestretch} {1.3}

\maketitle
\begin{abstract}
The situation connected with a possibility of CPT violation in neutrino
sector is considered. 
    
\end{abstract}

CPT theorem \cite{lude,pais,paul} states: the local theory of a quantum
field, invariant concerning Lorentz-rotations and including a usual causal
commutativity or an anticommutativity of operators of a field, is always
invariant relative products of CPT transformations.

From CPT theorem, the equality of matrix elements of processes P and CPTP
implies. And CPTP process turns out from P replacement of all particles
by antiparticles, all spins on inverse and permutation of
initial and final conditions.
In particular, from CPT theorem equalities of masses and life times, and also
difference only in a sign of the magnetic moments of particles and
antiparticles follow.

Let us notice that the theory of mixing of a neutrino  \cite{bile,grib,pont}
is essentially nonstationary. The born beam of a neutrino is described not
by a stationary state, as in traditional theories of strong, electromagnetic
and weak interactions, and superposition of stationary conditions:
\begin{eqnarray}
\label{eq:MM1}
\mid\nu_e(0)>=c_1\mid\nu_1>+s_1c_3\mid\nu_2>+s_1s_3\mid\nu_3>, 
\end{eqnarray}
where $c_i=cos\theta_i$, $s_i=sin\theta_i$, $\theta_i$ - angles of mixing
matrix.

Condition evolution in time is defined by corresponding eigenvalues of an
operator of energy at the fixed impulse $p$:
\begin{eqnarray}
\label{eq:MM2}
E^2_i=p^2+m^2_i, 
\end{eqnarray}
or
\begin{eqnarray}
\label{eq:MM3}
\mid\nu_e(t)>=c_1e^{-iE_1t}\mid\nu_1>+
s_1c_3e^{-iE_2t}\mid\nu_2>+s_1s_3e^{-iE_3t}\mid\nu_3>. 
\end{eqnarray}

The born beam has probability of a survival as oscillates owing to a
difference of masses
$m_1, m_2, m_3$. In model of two-neutrino mixing the probability of a survival
can be noted in terms of parametre of mixing
$sin^2 2\theta$ and a difference of quadrats of masses $\delta 
m^2_{\nu}=\mid m^2_2-m^2_1\mid$,
properly: 
\begin{eqnarray}
\label{eq:MM4}
P_{\nu_e \rightarrow \nu_e}=1-sin^2 2\theta_{\nu}
sin^2\Bigl(\frac{\delta m^2_{\nu}L}{4E}\Bigr), 
\end{eqnarray}
where $L$ - distance from a source to the detector.

If CPT theorem is fulfilled in neutrino sector, similar (\ref{eq:MM4})
relation can be fair and for  $\bar{\nu}$ \cite{giun}:
\begin{eqnarray}
\label{eq:MM5}
P_{\bar{\nu}_e \rightarrow \bar{\nu}_e}=1-sin^2 2\theta_{\bar{\nu}}
sin^2\Bigl(\frac{\delta m^2_{\bar{\nu}}L}{4E}\Bigr). 
\end{eqnarray}

Formulas (\ref{eq:MM4}) and (\ref{eq:MM5}) should be fulfilled at the same
distances from a source $L$ and at identical energies of neutrino and
antineutrino $E$.

In \cite{giun} check of correspondence of probabilities of the disappearance of
electron neutrino
$P_{\nu_e \rightarrow \nu_e}$ (\ref{eq:MM4}) and antineutrino
$P_{\bar{\nu}_e \rightarrow \bar{\nu}_e}$ (\ref{eq:MM5}) was spent.
Data of the radioactive source neutrino experiments
$^{51}Cr$ of collaboration Gallex and $^{51}Cr$, $^{37}Ar$ of collaboration
Sage presented in \cite{giunt} was thus used.
And also were used reactor antineutrino disappearance experiments of
collaborations
Bugey \cite{achk} and Chooz \cite{apol}.

For calculation of parametres $\delta m^2_{\nu}$, $sin^2 2\theta_{\nu}$ and
$\delta m^2_{\bar{\nu}}$, $sin^2 2\theta_{\bar{\nu}}$ the maximum
likelihood method was used. Also asymmetries for masses and mixing angles are 
calculated:
\begin{eqnarray}
\label{eq:MM6}
A^{CPT}_{\delta m^2}&=&\delta m^2_{\nu}-\delta m^2_{\bar{\nu}},\nonumber\\
A^{CPT}_{sin^2 2\theta}&=&sin^2 2\theta_{\nu}-sin^2 2\theta_{\bar{\nu}}. 
\end{eqnarray}

The best-fit values of the asymmetries with $\chi^2_{min}$ are
\cite{giun}:
\begin{eqnarray}
\label{eq:MM7}
A^{CPT}_{sin^2 2\theta}=0.42,~~A^{CPT}_{\delta m^2}=0.37 eV^2.  
\end{eqnarray}
Authors \cite{giun} consider there are indications on CPT violation in
disappearance experiments of electron neutrinos and antineutrinos by
confronting the neutrino data and the antineutrino data.

However there are doubts in it considering essential unhomogeneity of
experiments compared in
\cite{giun} of Gallex-Sage and reactor antineutrino disappearance experiments
on statistics and errors.
For Gallex and Sage are available on pair experiments  wich essentially differ
from each other:
\begin{eqnarray}
\label{eq:MM8}
(Gallex)~Cr1~P_{\nu_e \rightarrow \nu_e}&=&1.0 \pm 0.10;~~~
Cr2~P_{\nu_e \rightarrow \nu_e}=0.81 \pm 0.10, \nonumber\\
(Sage)~^{51}Cr~P_{\nu_e \rightarrow \nu_e}&=&0.95 \pm 0.12;~
^{37}Ar~P_{\nu_e \rightarrow \nu_e}=0.79 \pm 0.10.
\end{eqnarray}
Such scatter of results of experiments testifies either to game to the
statistican or about systematic shift.

In the analysis \cite{giun} results of reactor experiments were used:
\begin{eqnarray}
\label{eq:MM9}
(Chooz)~P_{\bar{\nu}_e \rightarrow \bar{\nu}_e}=1.01\pm0.04,\nonumber\\
(Bugey)~P_{\bar{\nu}_e \rightarrow \bar{\nu}_e}=1.0\pm0.035,
\end{eqnarray}
that is results not displaced from unit. Let us notice, that if experiment
Bugey
was with short flying bases as well as the experiments Gallex-Sage that
experiment Chooz was with intermediate base
$\sim$ 1 km.

Averages on a harmonics of $\delta m^2$ and $sin^2 2\theta$ in \cite{giun}
have essential uncertainty. Therefore is the most advisable
check of CPT invariance at definition on experiments of the ratio of
$P_{\nu_e \rightarrow \nu_e}/P_{\bar{\nu}_e \rightarrow \bar{\nu}_e}$ with 
an adequate accurace. And experiments should be fulfilled at the same baseline
$L$ both identical energies of a neutrino and antineutrino $E$ and to be
homogeneous for statistics and erros.

In \cite{acer} the weighted average for results of definition of
$P_{\nu_e \rightarrow \nu_e}$ in four experiments
(\ref{eq:MM8}) is used. There is problem on correctness of association of
results of pair experiments of
$Cr1$ and $Cr2$ and also pair of $^{51}Cr$ and $^{37}Ar$.

\begin{figure*}[hb]
\begin{center}
\epsfig{file=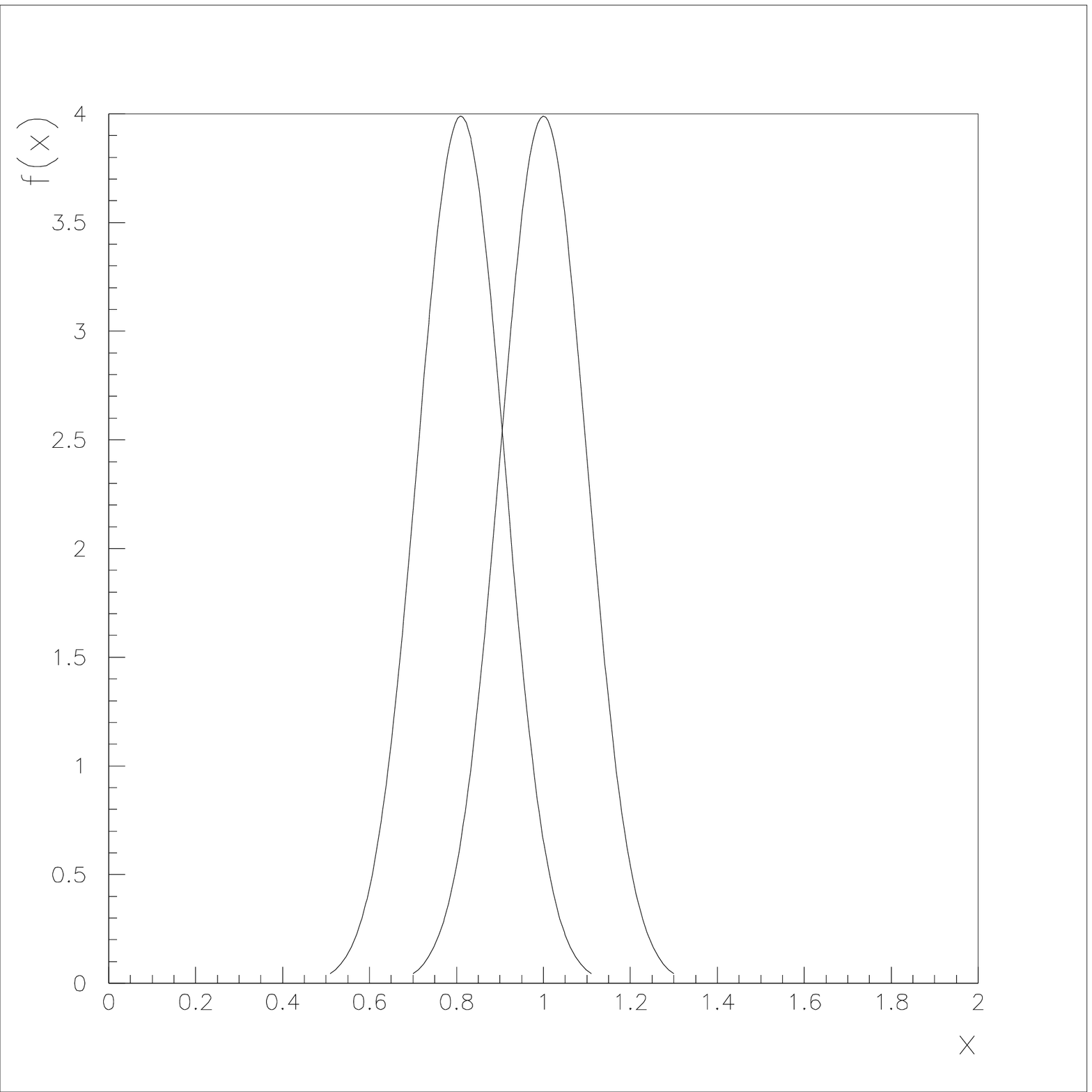,width=14.cm}
\end{center}
\caption{}
\label{fig:fi14}
\end{figure*}

We spend simple check on a homogeneity of experiments $Cr1$ and $Cr2$.
Assuming normal distribution with experimental values of an average and
dispersion for
$P_{\nu_e \rightarrow \nu_e}$ (Fig.\ref{fig:fi14}), it is received for product
of probability value 34\%. That already guards. If the difference of averages
in experiments $Cr1$ and $Cr2$ was up to level $1\sigma$ that product of
probability would be  61\%. That is association of results is more justified to
pair experiments
$Cr1$ and $^{51}Cr$.

\begin{figure*}[hb]
\begin{center}
\epsfig{file=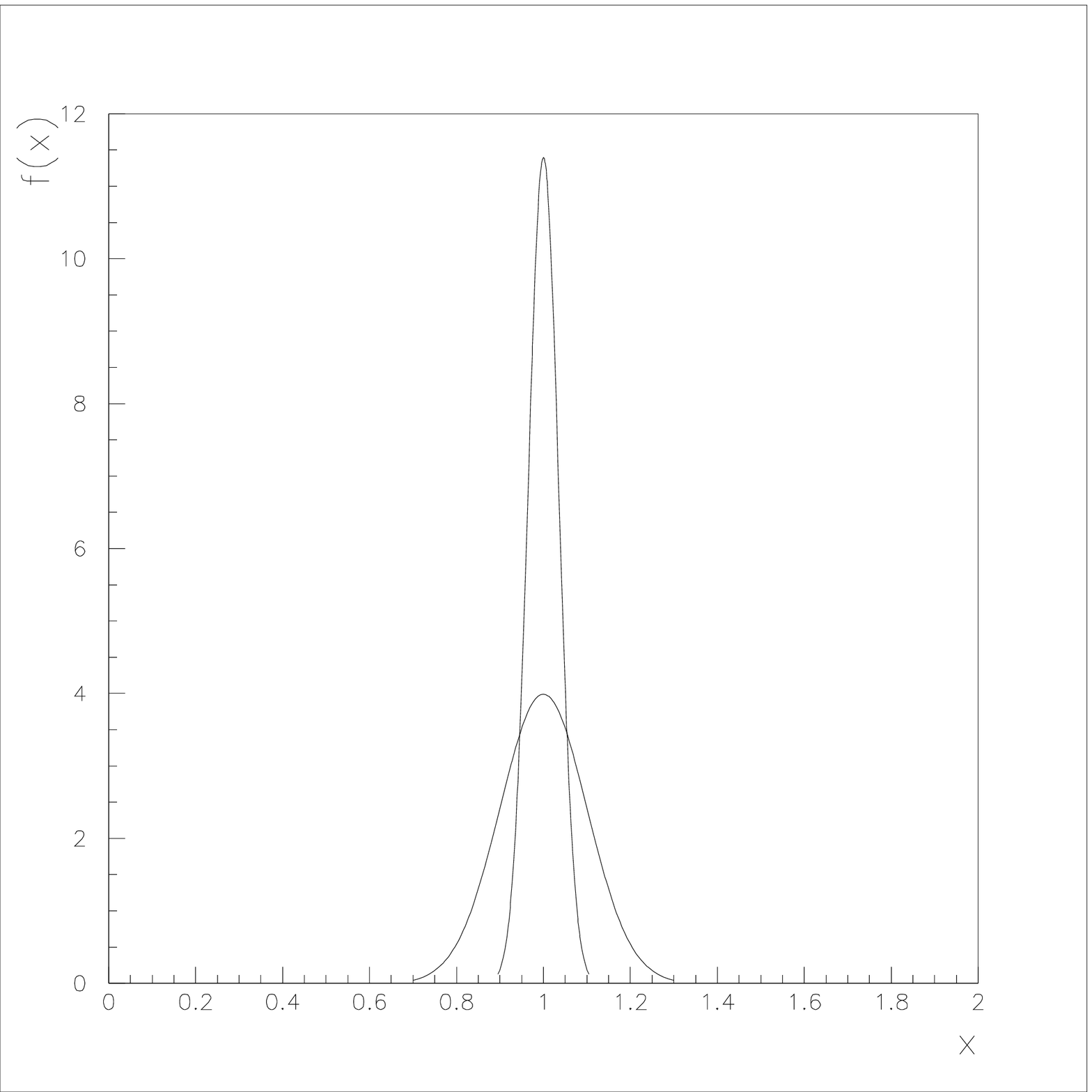,width=14.cm}
\end{center}
\caption{}
\label{fig:fi15}
\end{figure*}

In the same way it is possible to check up on a homogeneity results of
experiment of
$Cr1$ and experiment of $Bugey$ for an antineutrino  that already
reflects check on CPT invariance
(Fig.\ref{fig:fi15}). Also we receive for product of probability value 53\%.
Experiment of $Choos$ does not correspond to a principle of identical baseline
for check of CPT invariance.

Probably to unite pair of experiments $Cr1$ and $^{51}Cr$ and to receive a
weighted average for this pair:
\begin{eqnarray}
\label{eq:MM10}
P_{\nu_e \rightarrow \nu_e}=0.975\pm0.078.
\end{eqnarray}
Product of probability for this pair (\ref{eq:MM10}) and result of $Bugey$
(\ref{eq:MM9}) equals 61\%.

For homogeneous experiments of $Cr2$ and $^{37}Ar$ the weighted average
equals:
\begin{eqnarray}
\label{eq:MM11}
P_{\nu_e \rightarrow \nu_e}=0.80\pm0.071.
\end{eqnarray}
Product of probability for this pair of experiments (\ref{eq:MM11})
and experiment of $Bugey$
(\ref{eq:MM9}) equals 5\%.

So we see that experimental data available now
on disappearance of electron neutrino
with short-baseline are unsatisfactory for a solution of the problem on CPT
invariance in neutrino sector.
The possible solution of this problem is connected with the furure experiments
at accelerators.

At last we will notise that in \cite{giun,malt} approach it is impossible to
define errors in average $\delta m^2$ and $sin^2 2\theta$. Therefore in
\cite{giun} asymmetries (\ref{eq:MM7}) for masses and angles of mixing
without their errors are presented.
The confidence levels specified in \cite{giun} concern to
$P_{\nu_e \rightarrow \nu_e}$ and $P_{\bar{\nu}_e \rightarrow \bar{\nu}_e}$.
Also do not give us the information on uncertainty of asymmetries.
  
In summary it is necessary to tell about importance of check of
CPT invariance independent on model in neutrino sector, including from the
model of an oscillation of a neutrino accepted now.
Definition of the ratio
$P_{\nu_e \rightarrow \nu_e}/P_{\bar{\nu}_e \rightarrow \bar{\nu}_e}$ with
an adequate accuracy can be such check.

In this direction experiment of Collaboration MINOS (Main Injector Neutrino
oscillation Search) with long-baseline 734 km is perspective
\cite{mich,adam}. MINOS employs two detectors to significantly reduce the
effect that systematic uncertainties associated with the neutrino flux have
upon the $\nu_{\mu}$ and $\bar{\nu}_{\mu}$ disappearance measurement. From
recent results of MINOS follows that there is a certain difference between the
oscillation parameters for $\nu_{\mu}$ and $\bar{\nu}_{\mu}$ \cite{vahl}. At
90\% confidence level, it reports that:
\begin{eqnarray}
\label{eq:MM12}
|\delta m^2_{32}| = 2.35^{+0.11}_{-0.08} \times 10^{-3} ~eV^2,
\end{eqnarray}
\begin{eqnarray}
\label{eq:MM13}
|\delta \bar{m}^2_{32}| = 3.36^{+0.45}_{-0.40} \times 10^{-3} ~eV^2.
\end{eqnarray}
Together with $sin^2(2\theta_{23})>0.91$ и $sin^2(2\bar{\theta}_{23})=
0.86 \pm 0.11$.

Let us notice that on the long-baseline the effect of localisation of admissible
space of the oscillation parameters takes place. That has allowed a
goodnes-of-fit method similar in \cite{malt} to receive errors for
$\delta m^2_{32}$ (\ref{eq:MM12},\ref{eq:MM13}) and intervals for
$sin^2(2\theta_{23})$. And from short-baseline experiments calculated in
\cite{giun} asymmetries (\ref{eq:MM6}) have no corresponding errors. Therefore
results of evaluation of asymmetries (\ref{eq:MM7}) do not allow to draw any
conclusion on CPT violation. To be convinced of it, it is enough to look at
Figs.1,3 from \cite{acer} and on Fig.37 from \cite{adam1}.

So search of CPT violation in neutrino sector with definition of oscillation
parameters with short-baseline is represented unpromising. That confirms our
conclusion that definition on the ratio
$P_{\nu_e \rightarrow \nu_e}/P_{\bar{\nu}_e \rightarrow \bar{\nu}_e}$ with a
sufficient statistic can be check of CPT invariance.

By the way in \cite{vahl} along with research of oscillation parameters of
atmospheric neutrino and antineutrino the ratio of final flaxes of a neutrino
and an antineutrino normalised on Monte Carlo is presented also:
 \begin{eqnarray}
\label{eq:MM14}
P^{data}_{\bar{\nu}/\nu}\bigl/P^{MC}_{\bar{\nu}/\nu} =
1.04^{+0.11}_{-0.10} \pm 0.10.
\end{eqnarray}
Whence difference of thys ratio from unit within an error is visible that and
is impossible to speak about CPT violation. There are foundation for research
prologation in this direction.

\newpage
\clearpage

\newpage
\clearpage

\newpage
\clearpage

\begin{center}
Figure captions
\end{center}

Fig.~1. Check on a homogeneity of results of experiments of $Cr1$ and $Cr2$.

Fig.~2. Check on a homogeneity of results of experiment $Cr1$ and
        experiment of $Bugey$.

\end{document}